\documentclass[aps,prd,superscriptaddress,floatfix,showpacs,showkeys,preprintnumbers]{revtex4}
\usepackage{pdfsync}
\usepackage{mathptmx}
\usepackage[utf8]{inputenc}
\usepackage[T1]{fontenc}
\usepackage{slashed}
\usepackage{times}
\usepackage{amssymb,amsfonts,amsmath,amsthm}
\usepackage{dsfont,bbm}
\usepackage{dcolumn}
\usepackage{epsf}
\usepackage{graphicx}
\usepackage[caption=false]{subfig}
\usepackage{dsfont}
\usepackage{slashed}
\usepackage[active]{srcltx}
\usepackage[usenames]{color}

\begin{document}

\preprint{\it\small Contribution to the Special Issue “QCD and Hadron Structure”
devoted to the memory of Prof. G.V.~ Efimov}

\title{Estimation of nucleon D-term in QCD}

\author{I.~V.~ Anikin}
\email{anikin@theor.jinr.ru}
\affiliation{Bogoliubov Laboratory of Theoretical Physics, JINR,
             141980 Dubna, Russia}

\begin{abstract}
Using the light-cone sum rules at leading order, we present an approach to perform
the preliminary upper estimation for the nucleon gravitational form factor $\mathds{D}(t)$ ($D$-term contribution).
Comparison with the experimental data and with the results of different models is discussed.
\end{abstract}
\date{\today}
\maketitle

\section{Introduction}

It is well-known that the hadron matrix element of energy-momentum tensor (EMT) can provide information
on fundamental characteristics of particles such as mass and spin \cite{Polyakov:2002yz, Polyakov:2018zvc, Polyakov:2018exb, Belitsky:2005qn, Diehl:2003ny}.
One of the EMT form factors has been related to the $D$-term \cite{Kivel:2000fg} which can be called
as the last unknown fundamental hadron characteristic determining the spatial deformations as well as
defining the mechanical properties of hadrons \cite{Polyakov:2002yz, Polyakov:2018zvc, Polyakov:2018exb}.
The analogy of $D$-term with the vacuum cosmological constant has been studed in \cite{Teryaev:2013qba}.
The $D$-term has also been calculated in the framework of the dispersion relation approach where the good agreement with
the chiral quark-soliton model has been achieved \cite{Pasquini:2014vua}.
Recently, the results of \cite{Polyakov:2018zvc} have been extended to the different
frames where the nucleon has the non-vanishing average momentum \cite{Lorce:2018egm}.

Since the energy-momentum tensor is the interplay between the gravitation as a external field
and the matter fields (in full analogy with the gauge fields which interact with particles
through the corresponding electromagnetic current), we adopt
the light-cone sum rules (LCSRs) developed for the different nucleon electromagnetic form factors in
\cite{Anikin:2013aka, Anikin:2015ita, Anikin:2016teg}.

In the present paper, we develop the light-cone sum rules for the preliminary upper
estimation $M$ of the nucleon gravitational form factor $\mathds{D}(t)$, $|\mathds{D}(t)|\le M$.
Unfortunately, the full information on the gravitation form factor
$\mathds{D}(t)$  ($D$-term form factor) cannot be obtained within the LCSRs due to
the lack of the sum rules that use the minus and perpendicular light-cone projections of electromagnetic current.
To estimate $\mathds{D}(t)$ we study the valence quark contributions in nucleon that originate
from the leading collinear twist-$2$ combination of the electromagnetic current.
In order to approach the region of small $t$ , where the LCSRs approach is really useless,
we first approximate the form factor derived for the region of  large $t$ by
the appropriate multipole function and, then, we do the extrapolation of fitted form factor to
the region of small $t$.
Our final preliminary prediction demonstrates reasonably good agreement with the first experimental data \cite{Hagler:2007xi, Burkert:2018bqq} together with
the chiral quark-soliton and Skyrme model results for $D$-term contributions \cite{Polyakov:2018zvc}.

\section{Energy-momentum tensor}

For the quark contribution, the Belinfant\'{e} improved energy-momentum tensor
is nothing but the local geometrical twist-$2$ operator which reads
\begin{eqnarray}
\label{emt-1}
\Theta^{\mu\nu}_q(0)= \frac{i}{2} {\cal R}^{\mu\nu}_{\tau=2}
\end{eqnarray}
and it can be expressed through the non-local operator as (cf. \cite{Balitsky:1987bk})
\begin{eqnarray}
\label{emt-2}
-2i\,\Theta^{\mu\nu}_q(0)&=&\lim_{y\to 0}\frac{\partial}{\partial y_\nu}\, \int\limits_{0}^{1}du\,\frac{\partial}{\partial y_\mu}
\big[ \bar\psi(0)\, \hat y \,[0\,;\,uy]_A\, \psi(uy)  \big]
\nonumber\\
&-& \text{(trace)},
\end{eqnarray}
where $y=(y^+,y^-,\vec{\bf y}_\perp)$. From now on, we do not show the trace subtractions.
In the paper, we are restricted by the bilinear quark combinations with the spin projection $s_a=+1$.
Hence, we deal with the collinear twist-$2$ quark combination $[\bar\psi_+\, \psi_+]$ of ${\cal R}^{\mu\nu}_{\tau=2}$.
In this connection, we also introduce the local geometrical twist-2 operator which reads
\begin{eqnarray}
\label{tw-2-2}
&&\widetilde{\cal R}^{\mu\nu}_{\tau=2}=\lim_{y\to 0}\frac{\partial}{\partial y_\nu}\, \int\limits_{0}^{1}du\,\frac{\partial}{\partial y_\mu}
\big[ \bar\psi(0)\, y_\alpha\gamma^{\alpha,+} \,[0\,;\,uy]_A\, \psi(uy)  \big].
\end{eqnarray}
Notice that this operator actually differs from the operator ${\cal R}^{\mu\nu}_{\tau=2}$ but
$\widetilde{\cal R}^{+ +}_{\tau=2} = {\cal R}^{+ +}_{\tau=2}.$

Moreover, the operator ${\cal R}^{+-}_{\tau=2}$ can be written as
\begin{eqnarray}
\label{pmR}
{\cal R}^{+-}_{\tau=2}=&&
\frac{1}{2} \Big( \bar\psi(0)\, \gamma^{+}\vec{\cal D}^{-}\,\psi(0) +  \bar\psi(0)\, \gamma^{-}\vec{\cal D}^{+}\,\psi(0) \Big)
\nonumber\\
=&&\widetilde{\cal R}^{+-}_{\tau=2}  + \frac{1}{2} \bar\psi(0)\, \gamma^{-}\vec{\cal D}^{+}\,\psi(0),
\end{eqnarray}
where the first term with the quark combination involving $[\bar\psi_+\, \psi_+]$  defines $\widetilde{\cal R}^{+-}_{\tau=2}$ and
the second term with $[\bar\psi_-\, \psi_-]$ is kept intact because it is beyond the direct computations within our approach.
Despite, the $[\bar\psi_-\, \psi_-]$ contribution to the projection $\Theta^{+-}$, which we eventually need,
remains unavailable for the explicit LCSRs calculations, we however can still make some reasonable estimation for this kind of contribution.

\section{Gravitational form factors}

We now remind the standard definitions for the relevant gravitational form factors.
The quark contribution to the hadron matrix element of the Belinfant\'{e} improved energy-momentum tensor operator
reads (see, for instance, \cite{Polyakov:2018exb})
\begin{eqnarray}
\label{EMT-me-1}
&&\langle P^\prime | \Theta^{(q)}_{\mu\nu}(0) | P \rangle =
\bar N(P^\prime) \Big[ \mathds{A}(t)\frac{\overline{P}_\mu\overline{P}_\nu}{m_N} +
i\mathds{J}(t)\frac{\overline{P}_{\left\{\mu\right.}\sigma_{\left.\nu\right\}\Delta}}{m_N} +
\nonumber\\
&&
\mathds{D}(t)\frac{\Delta_\mu\Delta_\nu -g_{\mu\nu} \Delta^2}{4m_N} + g_{\mu\nu} m_N \overline{\mathds{C}}(t)
\Big] N(P),
\end{eqnarray}
where
\begin{eqnarray}
&&\mathds{J}(t)=\frac{1}{2} \Big( \mathds{A}(t) + \mathds{B}(t)\Big),\quad
a_{\left\{\mu\right.}b_{\left.\nu\right\}} = \frac{1}{2} \big( a_\mu b_\nu + a_\nu b_\mu\big), \quad
\nonumber\\
&&
\overline{P}=\frac{1}{2}\big( P^\prime + P\big), \quad \Delta=P^\prime - P, \quad \Delta^2=-t.
\end{eqnarray}
Notice that the Belinfant\'{e} improved energy-momentum or angular momentum tensor includes the contribution from the spin momentum tensor.
As usual, the hadron momenta can be expanded as \cite{Anikin:2013aka}
\begin{eqnarray}
P=p+\frac{m^2_N}{2}n, \quad \Delta=P\cdot \Delta\, n + \Delta_\perp.
\end{eqnarray}

To calculate the form factor $\mathds{D}(t)$ we have to concentrate on a projection ${\cal R}^{+-}_{\tau=2}$.
Since our exact computations are restricted by the consideration of $[\bar\psi_+ \, \psi_+]$ combination,
we are dealing with the corresponding
projections of $\widetilde{\cal R}^{\mu\nu}_{\tau=2}$ rather than the projections of ${\cal R}^{\mu\nu}_{\tau=2}$.
With respect to the $[\bar\psi_- \, \psi_-]$ combination, we propose a reliable recipe for estimations of this contribution.

In analogy with the nucleon electromagnetic form factors \cite{Anikin:2013aka, Anikin:2015ita, Anikin:2016teg},
we define the amplitude which corresponds to the hadron matrix element of the energy-momentum tensor as
\begin{eqnarray}
\label{gravi-amp-1}
&&\hspace{-0.4cm}T^{\mu\nu}_{[\bar\psi_+ \psi_+]}(P,\Delta)\equiv\langle P^\prime| \widetilde{\cal R}^{\mu\nu}_{\tau=2} | P\rangle
=\lim_{y\to 0}\frac{\partial}{\partial y_\nu}\, \int\limits_{0}^{1}du\,\frac{\partial}{\partial w_\mu}
\int(d^4z) e^{-i\Delta\cdot z} \langle 0| T \eta(0)
\nonumber\\
&&\hspace{-0.4cm}
\times\big[ \bar\psi(w+z)\, w^\alpha\gamma^{\alpha,+}\,
[w+z\,;\,-w+z]_A\, \psi(-w+z) \big]| P\rangle
\end{eqnarray}
where $w=uy$. $\eta(0)$ stands for the Ioffe interpolation current defined as
\begin{eqnarray}
\label{Ioffe}
\eta(0)=\varepsilon^{ijk}\big[ u^i(0) C \gamma_\alpha u^j(0)\big] \gamma_5 \gamma_\alpha d^k(0).
\end{eqnarray}
Below, the colour indices $i,j,k,..$ are omitted.

By the direct calculation of (\ref{gravi-amp-1}), using the distribution amplitude parametrization
(see for example \cite{Anikin:2013aka}), one concludes
that in order to extend the preceding calculations performed for
electromagnetic form factors to the case of gravitational form factors
we have to weight the electromagnetic form factors with the certain tensor structure.
So, the exact expression for the amplitude for $d$-quark contribution is given by
\begin{eqnarray}
\label{Adopt-2}
&&T^{\mu\nu\,(d)}_{[\bar\psi_+ \psi_+]}(P,\Delta) =
\\
&&
\frac{1}{8} \int {\cal D}x_i
\Big( 2p_{\mu} \big[ 2x_1 P + \Delta \big]_{\nu} + (\mu\leftrightarrow\nu) \Big)
T^{(d)\,+}_{em}(x_i; P,\Delta),
\nonumber
\end{eqnarray}
where
\begin{eqnarray}
\label{Adopt-2-2}
&&T^{(q)\,+}_{em}(x_i; P,\Delta) =
\\
&&
 \Big\{
m_N\, {\cal A}^{+\,(q)}_{em}(x_i; \Delta^2, P^{\prime\, 2}) +
\hat\Delta_\perp \,{\cal B}^{+\,(q)}_{em}(x_i; \Delta^2, P^{\prime\, 2})\Big\} N^+(P).
\nonumber
\end{eqnarray}
and ${\cal A}^{+\,(q)}_{em}$, ${\cal B}^{+\,(q)}_{em}$ have been taken from \cite{Anikin:2013aka}.
In a similar manner we can write down the $u$-quark contribution.

From eqn.~(\ref{Adopt-2}), we conclude that the tensor structure of the amplitude (\ref{gravi-amp-1})
is entirely determined by the tensor
\begin{eqnarray}
2p_{\mu} \big[ 2x_i P + \Delta \big]_{\nu} + (\mu\leftrightarrow\nu).
\end{eqnarray}

Since the Ioffe current $\eta(0)$ has been used as the interpolation current, we focus on the
valence quark contributions to nucleon. Notice that all amplitudes and form factors should be understood as the objects
where the summation over $u$- and $d$-flavours have been implemented:
$T_{\mu\nu}=T^{(u)}_{\mu\nu}+T^{(d)}_{\mu\nu}$ etc.

\section{Estimation of the form factor $ \mathds{D}(t)$}

We are now going over to the plus-minus light-cone projection of the amplitude which is $T^{+-}(P,\Delta)$.
For the sake of shortness, we introduce the combination given by
\begin{eqnarray}
\label{F}
\mathds{F}(t)= 2m_N\Big(m_N\overline{\mathds{C}}(t) - \frac{\Delta^2_\perp}{4m_N}\mathds{D}(t) \Big).
\end{eqnarray}
For our goal, we first focus on the $T_1$-type (or $\mathcal{A}$-type) of amplitudes, we have the following relations
(for the LCSR notations, see \cite{Anikin:2013aka})
\begin{eqnarray}
\label{Pol-sr-1}
&&\hspace{-0.6cm}\frac{\lambda_1 m_N}{m^2_N - P^{\prime\,2}}
\Big(
m^2_N \mathds{A}(t) + \Big[ \frac{\Delta^2_\perp}{4} + P\cdot\Delta \Big]\mathds{A}(t)
+ \frac{\Delta^2_\perp}{4}\mathds{B}(t)
\nonumber\\
&&\hspace{-0.6cm}
 + \mathds{F}(t)
\Big) N^+(P)
=
m_N \Big(
\bar T^{+-}_{1\,[\bar\psi_+ \psi_+]} + \bar T^{+-}_{1\,[\bar\psi_- \psi_-]}
\Big) N^+(P)
\end{eqnarray}
where $\mathds{A}(t)$ and $\mathds{B}(t)$ can be expressed through the plus-plus
projection of the amplitudes, $\mathcal{A}^{++}$ and $\mathcal{B}^{++}$ respectively,
and
\begin{eqnarray}
\label{Pol-sr-1-2}
&&\hspace{-0.5cm}m_N \Big(
\bar T^{+-}_{1\,[\bar\psi_+ \psi_+]} + \bar T^{+-}_{1\,[\bar\psi_- \psi_-]}
\Big)=m_N \bar T^{+-}_{1\,[\bar\psi_- \psi_-]} +
\\
&&\hspace{-0.5cm}
\frac{m_N }{\pi}\int_{0}^{s_0} \frac{ds}{s-P^{\prime\,2}} \text{Im}\Big\{
\frac{m^2_N}{4} \mathcal{A}^{++}(s,t) + \frac{P\cdot\Delta}{4}\mathcal{A}^+_{em}(s,t)
\Big\}.
\nonumber
\end{eqnarray}
Notice that unlike the plus-plus light-cone projection, the eqn.~(\ref{Pol-sr-1}) do not form
the closed system of equations. Hence, there is no way to compute the form factors $\mathds{D}(t)$
directly. Instead, we shall try to estimate the form factor $\mathds{D}(t)$.

The next step is to perform the Borel transforms of eqn.~(\ref{Pol-sr-1}).
Afterwards, we derive the first representation for $\mathds{F}(t)$ which reads
\begin{eqnarray}
\label{pm-proj-1}
&&\mathds{F}(t)=T^{+-}_{1\,[\bar\psi_- \psi_-]} -
\int \hat d\mu(s)\Big\{ \frac{m^2_N}{4} \mathcal{A}^{++} -
\nonumber\\
&&
 \frac{\Delta^2_\perp}{4} \Big(
\frac{1}{2} \mathcal{A}^{++} + \mathcal{B}^{++} - \frac{1}{2} \mathcal{A}^+_{em}
\Big)
\Big\},
\end{eqnarray}
where
\begin{eqnarray}
\label{DR-short}
\int \hat d\mu(s)\big\{\mathcal{F} \big\}
=\frac{1}{\lambda_1\pi}\int_{0}^{s_0} ds\, e^{(m_N^2-s)/M^2}\, \text{Im} \big\{\mathcal{F}(s,t) \big\}.
\end{eqnarray}

In the same manner, working with the $T_4$-type (or $\mathcal{B}$-type) of amplitudes, we derive the
the second equivalent representation for $\mathds{F}(t)$ which is
\begin{eqnarray}
\label{pm-proj-2}
&&\mathds{F}(t)=2\,\bar T^{+-}_{4\,[\bar\psi_- \psi_-]}
- \int \hat d\mu(s)\Big\{ \frac{m^2_N}{2}\Big( \mathcal{A}^{++} - \mathcal{B}^{++}\Big)
-
\nonumber\\
&&
\frac{\Delta^2_\perp}{4} \Big(
2\mathcal{B}^{++} - \mathcal{B}^+_{em}
\Big) \Big\}.
\end{eqnarray}
Combining (\ref{pm-proj-1}) and (\ref{pm-proj-2}), we obtain the relation as
\begin{eqnarray}
\label{pm-proj-f}
R(t)= \frac{4}{\Delta^2_\perp}\,
\Big\{ 2 \bar T^{+-}_{4\,[\bar\psi_- \psi_-]} -  \bar T^{+-}_{1\,[\bar\psi_- \psi_-]} \Big\}
\end{eqnarray}
where, by introduction,
\begin{eqnarray}
\label{R2}
R(t)=
\int \hat d\mu(s) \Big\{
\frac{\mathcal{A}^{++} }{2} - \mathcal{B}^{++} \Big\}\Big( 1+ \frac{2m^2_N}{\Delta^2_\perp} \Big)
-
\int \hat d\mu(s) \Big\{ \frac{\mathcal{A}^{+}_{em}}{2} - \mathcal{B}^+_{em}\Big\}.
\end{eqnarray}
%
\begin{figure}[ht]
\vspace{0.3cm}
\includegraphics[width=0.35\textwidth]{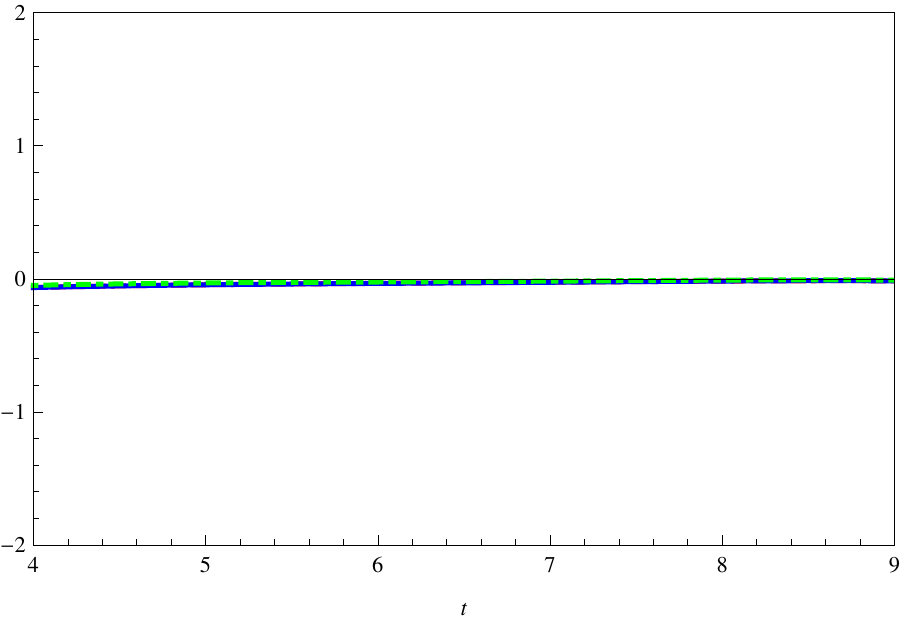}
  \caption{The functions $R(t)$.
  Notations: the solid blue line corresponds to ABO2; the dashed red line corresponds to ABO1; the dot-dashed green
  line corresponds to BLW \cite{Anikin:2013aka}.}
\label{Fig-R}
\end{figure}
%
The behavior of $R$ as functions of $t$ has been presented in Fig.~\ref{Fig-R}.
The numerical analysis, based on our work \cite{Anikin:2013aka}, shows that $R\approx 0$, therefore we are able to estimate the unknown LCSRs contributions as
\begin{eqnarray}
\label{Est}
 2 \lim_{t\to\infty}\bar T^{+-}_{4\,[\bar\psi_- \psi_-]} \approx \lim_{t\to\infty} \bar T^{+-}_{1\,[\bar\psi_- \psi_-]}.
\end{eqnarray}
As one can see in Fig.~\ref{Fig-R}, this estimation works with a good precision for $t\geq 4\, \text{GeV}^2$.

Further, since the amplitudes of $\mathcal{A}$- and $\mathcal{B}$-types are nothing but the invariant amplitudes of  $T_1$- and  $T_4$-types, having focused on $\bar T^{+-}_{1\,[\bar\psi_- \psi_-]}$ only, we impose that
\begin{eqnarray}
\label{Est-3-2}
\bar T^{+-}_{1\,[\bar\psi_- \psi_-]} \stackrel{t\to\infty}{\approx}
\int \hat d\mu(s) \Big\{ \frac{1}{2}\mathcal{A}^+_{em} -
\frac{1}{2}\mathcal{A}^{++}\Big( 1- \frac{2m^2_N}{t} \Big)  \Big\}.
\end{eqnarray}

We are now going over to the analysis of the $D$-term.
Using the QCD equations of motion,
we can express $\overline{\mathds{C}}(t)$ through the nucleon matrix elements of the quark-gluon operator
(for the quark contribution) or the gluon-gluon operator (for the gluon contribution) \cite{Tanaka:2018wea}.
Therefore, the calculation of $\overline{\mathds{C}}(t)$ is not available at the leading order at all.
Moreover, this form factor meets the condition $\sum_a \overline{\mathds{C}^a}(t)=0$ which shows that
the total (quark and gluon) energy-momentum tensor is conserved.
So, at the moment, the consideration of $\overline{\mathds{C}}(t)$ is not presented in the
present paper and is postponed to the forthcoming work.
Neglecting $\overline{\mathds{C}}$-term from
$\mathds{F}$-combination at the large $t$ and using the approximation (\ref{Est-3-2}), we estimate the
form factor $\mathds{D}(t)$ by (see, the Eqn.~(\ref{pm-proj-1}))
\begin{eqnarray}
\label{Dfull-Est}
\mathds{D}(t) \approx
 - \int \hat d\mu(s) \Big\{
\frac{1}{2}\mathcal{B}^{++} +
\frac{m^2_N}{t} \mathcal{A}^{++} \Big\}.
\end{eqnarray}
We emphasize that the form factor $\mathds{D}(t)$ is independent one.
Eqn.~(\ref{Dfull-Est}) shows that the estimated $\mathds{D}(t)$ as a function of $t$ behaves like
a combination of the form factors $\mathcal{A}^{++}$ and  $\mathcal{B}^{++}$.

\begin{figure}[ht]
\vspace{0.3cm}
\includegraphics[width=80mm]{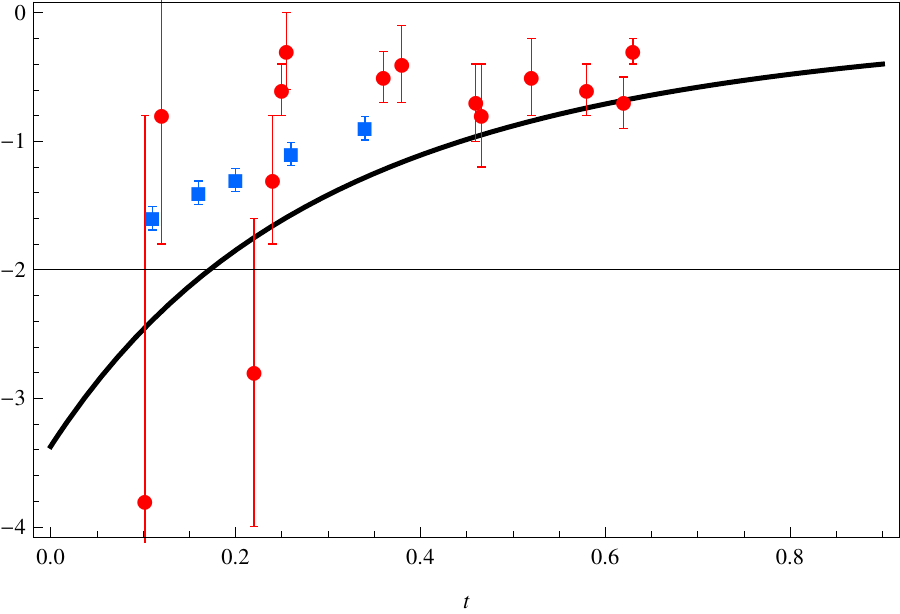}
  \caption{The preliminary upper estimation of $\frac{5}{4}\mathds{D}^{\text{fit}}(t)$ for the region of small $t$.
The experimental data on the nucleon $D$-term form factor:
the blue squares from the Jefferson Lab \cite{Burkert:2018bqq}, the red bullets from
the Lattice QCD \cite{Hagler:2007xi}.}
\label{Fig-D-fit}
\end{figure}

According to the strategy of \cite{Anikin:2013aka}, the gravitational form factor $\mathds{D}(t)$
is not reachable for a region of small $t$ within LCSRs approach.
However, we can do approximate fit for this form factor in the region of large $t$ and, then,
perform an extrapolation of the obtained fitting functions to the region of small $t$.
Finally, the preliminary upper estimation of $\mathds{D}(t)$ can be described by (cf. \cite{Tanaka:2018wea})
\begin{eqnarray}
\label{Dtot-fit}
&&\mathds{D}^{\text{fit}}(t) = \frac{D}{(1+e\, t)^n},
\nonumber\\
&&
D=-2.7, \quad e=0.96\,\text{GeV}^{-2}, \quad n= 3.43
\end{eqnarray}
In the region of small $t$, for all sets of LCSR-parametrizations, this approximating function matches, see Fig.~\ref{Fig-D-fit}, with the first experimental data
\cite{Burkert:2018bqq,Hagler:2007xi}
together with the results of the chiral quark-soliton and Skyrme models presented in \cite{Polyakov:2018zvc,Goeke:2007fp,Goeke:2007fq,Cebulla:2007ei}.
Our function $\mathds{D}^{\text{fit}}(t)$ lies below than the quark-soliton prediction and above than the result of Skyrme
model.

%
%

\section{Mechanical properties of nucleon}

Having calculated the form factor $\mathds{D}(t)$, one computes the pressure in the
center of nucleon and the hadron mechanical radius \cite{Polyakov:2018zvc}. We have
\begin{eqnarray}
\label{Pres}
p_0=-\frac{1}{24\pi^2 m_N} \int^{\infty}_{0} dt t \sqrt{t} \mathds{D}^{\text{fit}}(t)
= 0.79 \,\text{GeV} \text{fm}^{-3}
\end{eqnarray}
for the pressure $ [\text{cf.}\,\, p_0^{\text{\cite{Polyakov:2018zvc}}}=0.23 \,\text{GeV} \text{fm}^{-3}]$,
and
\begin{eqnarray}
\label{Rmech}
\langle r^2\rangle_{\text{mech}}=6D \Big[ \int^{\infty}_{0} dt \mathds{D}^{\text{fit}}(t)\Big]^{-1}
= 13.97 \,\text{GeV}^{-2}
\end{eqnarray}
for the mechanical radius.

\section{Conclusions}

The main challenge in the QCD description of any form factors is the calculation of soft contributions which
correspond to the Feynman mechanism to transfer the rather large momentum.
As demonstrated in many papers (see, for example, \cite{Balitsky:1986st, Balitsky:1989ry, Braun:2006hz, Braun:2001tj})
the LCSRs approach is attractive due to the fact that the soft contributions are calculated in terms of the same
DAs that the pQCD calculations include.
Thus, the LCSRs can be positioned as one of the most direct relations of the different kind of form
factors and hadron DAs that is available at present.

In the paper, we have presented the first preliminary estimation for the valence quark contributions to
the gravitational form factor $\mathds{D}(t)$ (the so-called $D$-term)
implemented due to the LCSRs techniques at the leading order.
To the region of  large $t$ where the LCSRs approach can be used,
we have observed that the estimated form factor $\mathds{D}$ as a function of $t$ has
the similar behaviour as the suitable combination of the well-known gravitational form factors $\mathds{A}(t)$ and $\mathds{B}(t)$.
Regarding the small $t$ region, where the first experimental data are available \cite{Hagler:2007xi, Burkert:2018bqq},
the estimated $\mathds{D}(t)$ can be approximated by the fitting multipole function
$\mathds{D}^{\text{fit}}(t)$. With the upper-estimated $\mathds{D}^{\text{fit}}(t)$, a few
quantities characterizing the so-called mechanical properties of the nucleon
have been calculated. We emphasize that
our result is consistent with the experimental data
as well as with the results of the chiral quark-soliton and Skyrme models presented in
\cite{Polyakov:2018zvc,Goeke:2007fp,Goeke:2007fq,Cebulla:2007ei}.



We are grateful to V.~Braun, M.~Deka, A.~Manashov, M.~Polyakov and P.~Schweitzer for useful discussions,
comments and remarks.



\end{document}